\documentstyle[12pt]{article}
\newcommand{\ox}{\otimes}
\newcommand{\gs}{\gamma}
\newcommand{\gl}{\Gamma}
\newcommand{\tgs}{\tilde{\gamma}}
\newcommand{\tgl}{\tilde{\Gamma}}
\newcommand{\w}{{\bf W}}

\newcommand{\wo}{\w_0}

\newcommand{\kij}{k_{ij}}

\newcommand{\tkij}{\tilde{k}_{ij}}

\newcommand{\na}{N_{\alpha \w}}

\newcommand{\h}{\frac{1}{2}}
\newcommand{\g}{\hspace{0.2 cm}}
\newcommand{\aw}{{\overline {\alpha \w}}}

\newcommand{\mod}{\mbox{\hspace{1 cm} mod(1)}}
\newcommand{\be}{\begin{equation}}
\newcommand{\ee}{\end{equation}}
\newcommand{\ba}{\begin{eqnarray}}
\newcommand{\ea}{\end{eqnarray}}
\newcommand{\ccc}{ {\mbox{\hspace{0.2 cm}}}\left| {\mbox{\hspace{0.2 cm}}} }
\newcommand{\lll}{\left[ {\mbox{\hspace{0.2 cm}}} }
\newcommand{\rrr}{{\mbox{\hspace{0.2 cm}}}\right.\right] }

\newcommand{\bb}{\tilde{b}}
\newcommand{\dd}{\tilde{d}}
\newcommand{\vv}{\tilde{v}}

\newcommand{\com}{}
\title{Classification of (2,2) Compactifications
       by Free Fermions 2}
\author{S.~A.~Abel$^{1}$ and C.~M.~A~Scheich$^{2}$\thanks{
Supported by the European Community, Science Program}
\vspace{0.3cm}
\\$^{1}$Rutherford Appleton Laboratory
\\Chilton, Didcot
\\Oxon OX11 0QX
\\England
\\ \vspace{0.3cm}
\\$^{2}$Departamento de Fisica Te\'orica, C-XI
\\Universidad Aut\'onoma de Madrid
\\28049 Madrid
\\Spain}

\begin{document}

\small

\maketitle

\title{Abstract}

\begin{abstract}
We present a classification of (2,2) free field compactifations with one twist
in which only 95 distinct models (generations and antigenerations)
are found. Models with three generations and no antigenerations are
given.
\end{abstract}
\vspace{-17.5cm}
\begin{flushright} {FTUAM-10/94}\end{flushright}

\newpage

Classification of spectra of different string compactifications  always serves
a twofold aim. On the one hand one is searching for realistic models with three
generations and as few antigenerations as possible. On the other hand one would
like   to get an overview of ``what a certain compactification scheme
contains'', especially in comparison to other schemes. Classification has
initiated enormous progress in understanding of the underlying relations
between different schemes.  In particular, analysis of the $(2,2)$ spectra
yielded by minimal $N=2$ models \cite{G88} led to the observation that they
realize Calabi-Yau manifolds at specific points in their moduli spaces. \hfill
\break Of the known string compactification schemes there are two of which
hardly anything is known concerning their content in terms of vacuum zero
modes: Lattice compactifications \cite{who}  and compactifications by
free fermions, also called fermionic strings \cite{ABC87,KLT87}. Here we will
be concerned with the  second case.\hfill \break Classification in this case
has been hampered by the huge number of possibilities for boundary conditions
for the fermions \cite{sene}. In most cases fermions with only periodic and
antiperiodic boundary conditions were used, which implied the need to introduce
several sets of boundary conditions to create viable models. This strategy was
adopted in most of the subsequent literature. \hfill \break Instead, in a
previous paper \cite{us} we proposed the opposite approach,  which is to
classify models with very general boundary conditions and a  minimum number of
different sets. Since it is believed that the main features of the vacua  are
already evident in the possible $(2,2)$ models,  we choose to concentrate on
them. Furthermore we impose left--right symmetry in anticipation of a possible
geometric interpretation.  Only with these restrictions is the classification
possible.

In this spirit, we gave a prescription for generating all possible left--right
symmetric $(2,2)$ models  in the fermionic formulation. Our aim in that work
was to make some general observations regarding the nature of fermionic string,
and its relation to other compactifications. We stressed that the $(2,2)$
structure is realized on the spectrum (e.g. implying space-time supersymmetry,
exceptional gauge groups, the existence of moduli), but that the explicit
formulation of the algebra in terms of general complex fermions is still
unknown. This situation is reminiscent of the one for Calabi-Yau
compactifications. Confining ourselves to $D=6$ and $D=8$ dimensions, we
discovered that there is a considerable  overlap with orbifolds and torus
compactifications, but that there exist many models in the fermionic
formulation which do not belong to any orbifold or known smooth manifold.
Specifically, in $D=8$ there exist only the known tori and no orbifolds. In
$D=6$ dimensions, we found 37 models of which 6 belong to the two-torus $T^2$
and  only 4 had a generation number which could possibly correspond to
orbifolds or the Calabi-Yau manifold K3.  (In $D=6$, the generation number is
related to the Hodge numbers by $n_+-n_-= h^{11}$ \cite{HM87}. All orbifold
models and the $K3$ manifold have 10 generations.) Only one of these models
could directly be bosonized, namely into the $Z_2$ orbifold. The other models
show very similar spectra, therefore suggesting highly nontrivial identities
similar to those proposed in ref.\cite{PS89} between the partition functions.

In this work, we extend the analysis to the case of $D=4$. Here we already have
the examples of equivalence between the fermionic models and the $Z_2$, $Z_4$
and $Z_8$ orbifolds by the already mentioned types of partition function
identities \cite{PS89,BDL90}.  The exact overlap between the two schemes
remains contentious however, and it is important to note that the fermionic
versions of these orbifolds were established using theta-function identities
and not direct bosonisation which indeed does not appear to be possible for the
$Z_4$ and $Z_8$ cases.

An additional aim here is to refute an assertion which often is made, namely
that there exists a unique way of generating three generation models which
involves a large set of boundary conditions.  In fact, extrapolating from the
$D=6$ case, one would naturally expect there to be many more than 37
left--right symmetric models in four dimensions. This makes the existence of a
unique theory unlikely (being in accordance with the Calabi-Yau and
Landau-Ginzburg schemes, in which there are also several three generation
models \cite{candelas,G88}). Our method for
generating left--right symmetric models was given in ref.\cite{us}, but for
completeness we shall briefly summarise our choice of vectors of boundary
conditions.

In our classification we will make use of the fact that $N=1$ space-time
supersymmetry is equivalent to $N=2$ world-sheet supersymmetry
\cite{dixon} (and the same is valid after the bosonic string map, turning
$N=1$ space-time supersymmetry into an  $E_{4+D/2}\otimes E_8$ gauge group).
Furthermore a model with  local $N=1$ space-time supersymmetry at the massless
level implies the existence of gravity supermultiplets. Since gravity couples
universally to all massless and massive states, it forces them all to appear in
supermultiplets. This implies $N=1$ space-time supersymmetry even at the
massive level, which by  the above theorem implies $N=2$ supersymmetry on the
world-sheet  (as does $E_{4+D/2}\otimes E_8$ after the heterotic string map).
\hfill \break We therefore shall restrict ourselves to left--right symmetric
$(1,1)$ models which are promoted into $(2,2)$ models in this manner.  Further
breaking of the gauge group by embeddings of twists  (e.g. Wilson-lines) should
then work in the usual way, and will not spoil the relevance of the
classification. We use the formulation of ref.\cite{KLT87}, and we stress that
we are restricting the analysis to only complex fermions. The internal degrees
of freedom then have phases associated with them $a_r,b_r,c_r$; $r=1,\cdots ,3$
which come in triplets for left and right movers fulfilling the constraint
\begin{equation}
a_r+b_r+c_r \g\in\g
 0,\frac{1}{2} \mod
\end{equation}
and therefore constituting a product of three $(1,1)$ models to start with.
\hfill \break Without loss of generality (see ref.\cite{us}) we choose the
first four vectors to be of the form,
\ba
\wo & = &
\lll (\h)\com (\h\com\h\com\h)^3
\ccc (\h\com\h\com\h)^3\com
(\h)^5\com(\h)^8 \rrr
\nonumber\\
\w_1 & = &
\lll (\h)\com (a^r_1\com b^r_1\com c^r_1)
\ccc (0\com 0\com 0)^3\com
(0)^5\com(0)^8 \rrr
\nonumber\\
\w_2 & = &
\lll (0)\com (0\com 0\com 0)^3
 \ccc (a^r_1\com b^r_1\com c^r_1)\com
(\h)^5\com(0)^8 \rrr
\nonumber\\
\w_3 & = &
\lll (0)\com (0\com 0\com 0)^3
\ccc (0\com 0\com 0)^3\com
(0)^5\com(\h)^8 \rrr.
\label{eq:4vectors}
\ea
The $\wo$ vector is needed to have a non-trivial modular invariant theory, and
to give the gravity multiplet. It implies the existence of Ramond and
Neveu-Schwarz sectors as in any string compactification. The $\w_1$ and $\w_2$
vectors respectively implement supersymmetry on the right movers and
exceptional gauge groups on  the left movers. Finally, in order to give a
second seperate $E_8^{\prime}$ factor we have the $\w_3$ vector. Thus we are
able to get copies of $N=2$ algebras on each side, establishing a $(2,2)$ model
\cite{us}.

The numerical survey of the spectra generated by the above vectors reveals
that, for any choice of $\mbox{($a^r_1$, $b^r_1$, $c^r_1$)}$, the theory
generated has the maximal $N=4$ supersymmetry (and so $E_8\ox E_8^{\prime}$
gauge group), and therefore corresponds to a torus compactification in the
usual sense. Typically one finds tori with enhanced symmetry. For example,
consider the choice $a_1^r=b_1^r=0$ and $c_1^r=\frac{1}{2}$ for all $r$. Direct
bosonisation (of the first two fermions in each triplet) shows that we do not
simply obtain a product  of three independent tori of radius $R=1/2$, since the
vector  $\tilde{\w_0}=\wo-\w_1-\w_2-\w_3$ relates them in a non-trivial way.
To go beyond torus compactification, we will need to add more vectors to break
down supersymmetry and gauge symmetry. Such additional  compactification
vectors may be either left--right symmetric,
\be
\w_4 =
\lll (0)\com (a^r_4\com b^r_4\com c^r_4) \ccc
(a^r_4\com b^r_4\com c^r_4) \com
(0)^5\com(0)^8 \rrr,
\ee
or may occur in left--right symmetric pairs,
\ba
\w_4 &=&
\lll (0)\com (a^r_4\com b^r_4\com c^r_4) \ccc
(a^r_5\com b^r_5\com c^r_5) \com
(0)^5\com(0)^8 \rrr \nonumber\\
\w_5 &=&
\lll (0)\com (a^r_5\com b^r_5\com c^r_5) \ccc
(a^r_4\com b^r_4\com c^r_4) \com
(0)^5\com(0)^8 \rrr,
\label{eq:doppel}
\ea
and so on. Usually it is assumed that only the first possibility  may allow the
interpretation of the model as a compactified variety  (e.g. in
ref.\cite{PS89}). However we emphasise that one should also  consider the
second possibility. This is similar to the case of the  comparison between
Calabi-Yau manifolds and compactifications by  products of $N=2$ models, where
the vacua of the latter are not always left--right symmetric.

For $N=1$, resp. $N=2$ the theories generated have the gauge group
\be
\label{G}
G=g\ox E_6\ox E_8^{\prime},
\ee
\be
G=g\ox E_7\ox E_8^{\prime},
\ee
where the first group, $g$ (which is of rank 8, resp. 7),  is some product of
low rank subgroups coming from the compactified degrees of freedom. In
ref.\cite{us} we found that with such a choice of vectors one should obtain all
possible left--right symmetric models, provided that one considers $\kij$
structure constants consistent with the preservation of modular invariance.
However this selection of vectors above is not sufficient to guarantee a (2,2)
compactification since we still have to choose the structure constants. A poor
choice of $\kij$ can spoil an ($N=2$) algebra by projecting out some of the
supersymmetry generators via the modular invariance conditions. This implies
the breaking of $N=1$ space-time supersymmetry and/or the exceptional group.
\hfill \break For any $(2,2)$ model there are always several choices of such
$k_{ij}$. E.g. they are fixing representations and antirepresentations.
\hfill \break In order to guarantee
a (2,2) model we need to impose a condition on the structure constants. We
usually do this by insisting that, given a gauge group $G$, the structure
constants are such that there are the required number of gravitino degrees of
freedom. A sufficient condition for this is \cite{us},
\be
\label{const}
\kij+\tkij=0 \mod ,
\ee
where the tilde implies the left--right reflected indices (for example
$\tilde{k}_{10}=k_{20}$, $\tilde{s}_1=s_2$ etc.). This always works because of
the chirality degrees of freedom of the gravitino and gaugino\footnote{We
confess that this is the correct version of Eq.(7) of ref.\cite{us}.}.
Using this
restriction, one only has to ensure that the gauge group has the structure $G$
above by the choice of $k_{ij}$.

We have examined $\sim 10^7$ possible models with  one symmetric twist vectors
upto order 20 and with the $\w_1$, $\w_2$ vectors containing fractions of
$\frac{1}{2}$ or 0 only, in most of the cases. Not taking the simplest
$\w_1$, $\w_2$ gives only a few additional models with low numbers of
generations and antigenerations. We shall discuss this point in more detail
below. The ``uncompactness'' of the fermionic string construction prohibits a
more complete classification than this, although we find that the number of new
models drops off very quickly as the number of twist vectors is increased as a
result of the more and more restrictive constraints for a modular invariant
theory. Concerning the increase of the order of the model, we checked a
further $10^5$ models upto order 40 and no new ones were found.
Therefore we believe that almost all possible spectra have been found.
We find approximately $10^4$ models of which $10^3$ have distinct spectra,
but most of them differ only in the number of singlets. Many of them
are related  as in the $N=2$ minimal models, where for instance in $D=6$ there
exist only two distinct models, namely $T^2$ and $K3$ \cite{G88}. Another
example is the case of $D=6$ compactification  by free fermions, where it was
found that there are two models corresponding to the $Z_2$  orbifold with such
enhanced symmetries \cite{us}. Beyond that one expects mirror symmetry
to be at work.

In this letter we shall only give the models with lowest order and  maximal
gauge group for each generation number. There are 95 distinct cases.  In table
1 we have displayed the internal part of the compactification vectors which can
achieve them in conjunction with the choice of vectors specified above.

Let us now discuss the relation to orbifolds. As was pointed out in
ref.\cite{PS89}, only $Z_N$, $Z_N\times Z_M$ orbifolds, where $N,M$ are powers
of 2, have any chance to be equivalent to fermionic strings given our current
knowledge about partition function identities. \hfill \break By directly
bosonising the $Z_2$ orbifold, one might expect it to have a compactification
vector
\[
\w_4 =
\lll (0)\com
\left(0\com\frac{ 1}{ 2}\com\frac{ 1}{ 2}\right)\com
\left(0\com\frac{ 1}{ 2} \com\frac{ 1}{ 2}\right)\com
\left(0\com 0\com 0\right) \ccc
\left(0\com\frac{ 1}{ 2} \com\frac{ 1}{ 2}\right)\com
\left(0 \com\frac{ 1}{ 2} \com\frac{ 1}{ 2}\right)\com
\left(0\com 0\com 0\right)\com
(0)^5\com (0)^8 \rrr .
\]
But here one should be careful, since as discussed above, our starting point
was a torus with enhanced symmetries due to the  vector $\tilde{\w}_0$
discussed above. Indeed the calculation shows that we have a model with six
generations of {\bf 56} representations of $E_7$, 96 singlets and  37
additional gauge bosons. Requiring a sectorwise equivalence of the partition
functions (as in ref.\cite{PS89}) one has to introduce the additional vector
\[
\w_5 =
\lll (0)\com
\left(\frac{ 1}{ 2} \com\frac{ 1}{ 2} \com 0\right) \com
\left(\frac{ 1}{ 2} \com\frac{ 1}{ 2} \com 0\right)\com
\left(0\com 0\com 0\right)\ccc
\left(\frac{ 1}{ 2} \com\frac{ 1}{ 2} \com 0\right)\com
\left(\frac{ 1}{ 2} \com\frac{ 1}{ 2} \com 0\right)\com
\left(0\com 0\com 0\right)\com
(0)^5\com (0)^8 \rrr .
\]
As expected, with this set of vectors we obtain the complete spectrum of the
$Z_2$ orbifold (10 generations and 80 singlets).  This is also apparent from
the fact that such a vector is needed to completely decouple one torus from the
internal part of the corresponding $D=6$ model. More specifically, we need to
break an initial, enhanced $SO(8)$ symmetry, down to $SO(4)\times SO(4)$. Using
this vector we recover, in addition to the $N=2$ models in the table, all  the
models of ref.\cite{us} with the obvious changes.\hfill \break

The non-singlet spectrum of the 27-3 version of the $Z_4$ orbifold (the singlet
numbers are not available in the literature, here we find 270 singlets and 20
additional gauge bosons) is generated by the first four vectors plus the vector
\[
\w_4 =
\lll (0)\com
\left(0\com\frac{ 1}{ 4} \com\frac{ 3}{ 4}\right) \com
\left(0\com\frac{ 1}{ 2}\com\frac{ 1}{ 2}\right)\com
\left(\frac{ 1}{ 2} \com\frac{ 3}{ 4} \com\frac{ 3}{ 4}\right)\ccc
\left(0\com\frac{ 1}{ 4} \com\frac{ 3}{ 4}\right)\com
\left(0\com\frac{ 1}{ 2} \com\frac{ 1}{ 2}\right)\com
\left(\frac{ 1}{ 2} \com\frac{ 3}{ 4} \com\frac{ 3}{ 4}\right)\com
(0)^5\com (0)^8 \rrr .
\]
Adding the $\w_5$ vector to the above gives the  31-7 version of the $Z_4$
orbifold (with 254 singlets and 12 additional  gauge bosons) \cite{EK93}. This
is in accordance with ref.\cite{PS89}, where the authors chose a slightly
different form of the superpartner of the  stress-energy tensor and slightly
different boundary conditions.  They found a 29-5 model (similar to a $Z_6$ or
$Z_{12}$ orbifold), which was turned into the 31-7 version of the $Z_4$
orbifold by adding the vector $\w_5$. \hfill \break The great majority of work
on fermionic strings has been based on the 27-3 left--right symmetric model
above. Traditionally this model is achieved (with exactly the same spectrum)
using a pair of symmetric compactification vectors \cite{far92}
\[
\w_4 =
\lll (0)\com
\left(0\com\frac{ 1}{ 2}\com\frac{ 1}{ 2}\right)\com
\left(0\com\frac{ 1}{ 2}\com\frac{ 1}{ 2}\right)\com
\left(0\com 0\com 0 \right) \ccc
\left(0\com\frac{ 1}{ 2}\com\frac{ 1}{ 2}\right)\com
\left(0\com\frac{ 1}{ 2}\com\frac{ 1}{ 2}\right)\com
\left(0\com 0\com 0 \right) \com
(0)^5\com (0)^8 \rrr
\]
\[
\w_5 =
\lll (0)\com
\left(\frac{1}{2}\com 0\com\frac{ 1}{ 2}\right)\com
\left(0\com 0\com 0 \right) \com
\left(0\com \frac{1}{2}\com\frac{ 1}{ 2}\right)\ccc
\left(\frac{1}{2}\com 0\com\frac{ 1}{ 2}\right)\com
\left(0\com 0\com 0 \right) \com
\left(0\com \frac{1}{2}\com\frac{ 1}{ 2}\right)\com
(0)^5\com (0)^8 \rrr .
\]
With this choice of vectors, a direct bosonisation exists along the lines of
ref.\cite{BDL90}. First we label the nine right moving internal  fermions by
\[
(\rho_1,\sigma_1,\psi_1)(\rho_2,\sigma_2,\psi_2)(\rho_3,\sigma_3,\psi_3)
\]
or in real fermions
\[
(\rho_1^{r1},\rho_1^{r2};\sigma_1^{r1},\sigma_1^{r2};\psi_1)
(\rho_2^{r1},\rho_2^{r2};\sigma_2^{r1},\sigma_2^{r2};\psi_2)
(\rho_3^{r1},\rho_3^{r2};\sigma_3^{r1},\sigma_3^{r2};\psi_3)
\]
and similarly the left movers. Then we split the triplets into the fermions
which have an odd phase under the supersymmetry vector $\w_1$ ($\psi_1$,
$\psi_2$, $\psi_3$) and the rest. The latter we wish correspond to complex
bosons ($z_1$, $z_2$, $z_3$). One defines the bosons as
\[
\frac{1}{\sqrt{2}}(\rho_i^{r1}+i\sigma_i^{r1})=
         :e^{iRez}:\ ;\ \
\frac{1}{\sqrt{2}}(\rho_i^{r2}+i\sigma_i^{r2})=
         :e^{iImz}:\ ,
\]
thus getting $Z_2$ twists on the bosonic coordinates. Obviously, for a
left-right symmetric model we need to do the same for the left movers. Thus we
may write down the action on the new coordinates ($\psi_i$, $z_i$) of various
combinations of compactification vectors,
\ba
\w_4       &:& \mbox{($\psi_1$, $z_1$), ($\psi_2$, $z_2$)}
              \rightarrow
  \mbox{($-\psi_1$, $-z_1$), ($-\psi_2$, $-z_2$)}\nonumber\\
\w_5       &:& \mbox{($\psi_1$, $z_1$), ($\psi_3$, $z_3$)}
              \rightarrow
  \mbox{($-\psi_1$, $-z_1+\pi+i\pi$), ($-\psi_3$, $-z_3$)}\nonumber\\
\w_4 + \w_5 &:& \mbox{($\psi_1$, $z_1$), ($\psi_2$, $z_2$)
                                         ($\psi_3$, $z_3$)}
              \rightarrow
  \mbox{($\psi_1$, $z_1+\pi+i\pi$), ($-\psi_2$, $-z_2$),
                                    ($-\psi_3$, $-z_3$)}.
\ea
It is easy to show that this may always be done if we only have phases of $\h$
or 0. On the other hand one could decide to take both real  components of a
complex field into a real boson. Then the situation is completely different,
since we never get twists - only shifts of the bosonic coordinates are created.
This therefore gives us a hint that twists in an orbifold may be reformulated
via the fermionic formulation as shifts. So we conclude that in the case of the
$Z_4$  orbifold we are only able to make the action of the $Z_2$ subroup
visible as twists, while the remainder is still hidden as shifts.

{}From table 1 we see that there are further similarities between spectra. But
now the discrete symmetries are sometimes completely different thus making any
conclusion difficult. We find models with the same spectra as the ($Z_4$),
$Z_8$, $Z_3\times Z_3$, or $Z_6\times Z_6$ orbifolds; $Z_4$, $Z_6$, $Z_{12}$
orbifolds; $Z_7$, $Z_8$ orbifolds; $Z_2 \times Z_6$ orbifolds. For a
comparison see ref.\cite{ibanez,EK93}.

Also from refs.\cite{ibanez,wir2} one finds no overlap with Gepner models
except the model no.83 in the table with five generations and one
antigeneration. This is similar to a Gepner model, namely the well studied
$3^5$ model \cite{G88,zoglin}\footnote{In the  $Z_5$ phase and $Z_5$
permutationally modded $3^5$ model one finds 5 generations, 1 antigeneration,
one additional  gauge boson and 42 singlets.}. However the discrete symmetries
are completely different for most of the cases
and also there is not the usual relation that models
are the same up to pairs of additional gauge bosons and singlets \cite{G88}.
\hfill \break Comparing the spectra to that of ref.\cite{coset}, one gets the
impression that the models studied here must be related to some varieties with
torsion.

Since there is a great deal of interest in three generation models and the
question of why we have just three generations, we should discuss a certain
peculiarity of our survey in detail. If one chooses the simplest set of $\w_i$
vectors (here having in mind a possible bosonisation as discussed above),
three generations occur quite naturally as the lowest possible number of
generations.
\\
Consider the $\wo $ sector in such a model where
{\bf 10} representations of $SO(10)$ always arise from
\be
b^{i}_{-\h} |0\rangle\ox \bb^{i}_{-\h} \bb_{-\h} |0\rangle
\ ; \
d^{i}_{-\h} |0\rangle\ox \dd^{i}_{-\h} \bb_{-\h} |0\rangle
\ee
excitations, where $i = (3,6,9)$. It is simple to show that these states always
satisfy the modular invariance conditions since they are symmetric in left and
right excitations. We still need to show that the {\bf 27} representations have
the same chirality, which we can do by examining the corresponding space-time
fermionic {\bf 16} states, which occur in the $\overline{\wo + \w_1 + \w_2}$
sector. The modular invariance conditions constrain their chirality;
\be
\begin{array}{ccccc}
\wo &:& \gl_5\gs_3\gs_6\gs_9 &=&
        \tgl_5\tgs_3\tgs_6\tgs_9 (-1)^{2(k_{01}+k_{02} + \h)} \nonumber\\
\w_1 &:& \gl_5\gs_3\gs_6\gs_9 &=&
          (-1)^{2(k_{11}+k_{12} +\h )} \nonumber\\
\w_2 &:& 1 &=&
        \tgl_5\tgs_3\tgs_6\tgs_9 (-1)^{2(k_{21}+k_{22})} \nonumber\\
\w_4 &:& \gs_3^{2W_4^3}\gs_6^{2W_4^6}\gs_9^{2W_4^9} &=&
        \tgs_3^{2W_4^3}\tgs_6^{2W_4^6}\tgs_9^{2W_4^9}
          (-1)^{2(k_{41}+k_{42})}
\end{array}
\ee
where we have labelled the internal degrees of freedom $1,\cdots, 9$. The first
condition is given by the $\w_1$ and $\w_2$ conditions via the structure
constant relations. Without loss of generality we can choose the structure
constants to be zero. Generically, the only solution to the $\w_4$ constraint
(which corresponds to the {\bf 10}) is $\gs_{i}=\tgs_{i}=\pm 1$, which gives
the three fermionic {\bf 16} with spin structure
$(\gs_3\gs_6\gs_9)=(++-),(+-+),(-++)$ and their antiparticles with
$(\gs_3\gs_6\gs_9)=(--+),(-+-),(+--)$ and in addition two chiralities of
gaugino with $(\gs_3\gs_6\gs_9)=(---),(+++)$. All three matter multiplets have
$\gl_5=-\tgl_5$ and thus the chirality of the {\bf 16} is the same in each {\bf
27}. (Alternatively we could have established this by examining their charges.)
\hfill \break Thus proving that at least three generations appear for the
simplest choice of $\w_1$ and $\w_2$ vectors.

Whilst the three generation models are of immediate interest for possible
phenomenological considerations\footnote{Not discussing here the question of
Wilson line breakings, which are still to be done to implement
reasonable gauge groups.},
the other ones seem to be not  so attractive at first glance. However studying
specific examples gives  the impression that it might also be possible to
promote those into models with three net generations (now with
antigenerations\footnote{A pattern which is
prefered by certain potentially viable schemes for a realistic
phenomenology \cite{ross}.}) by adding additional boundary vectors, naturally
leading to $(2,0)$ models. \hfill \break We shall demonstrate this by showing
two examples. The first model is initially a left--right symmetric 7-1 model
with the compactification vector
\be
\w_4 =
\lll (0)\com
\left(0\com\frac{ 1}{ 2}\com\frac{ 1}{ 2}\right)\com
\left(\frac{1}{3} \com\frac{3}{4} \com\frac{11}{12}\right)\com
\left(0\com\frac{5}{12} \com\frac{7}{12}\right) \ccc
\left(0\com\frac{ 1}{ 2} \com\frac{ 1}{ 2}\right)\com
\left(\frac{1}{3} \com\frac{3}{4} \com\frac{11}{12}\right)\com
\left(0\com\frac{5}{12} \com\frac{7}{12}\right)\com
(0)^5\com (0)^8 \rrr,
\ee
and will lead to a 4-1 model. Initially fermionic {\bf 16} representations come
from the  $\overline{\wo + \w_1 + \w_2}$ and  $\overline{\wo + \w_1 + \w_2 + 6
\w_4}$ sectors. For the first sector the  modular invariance projections are as
above, and give 3-1 generations, with  the following chiralities
\ba
\label{eq:chirality}
{\mbox{{\bf 16}}}    &:& (+++--++-),(++-+-+-+),(+-++--++)\nonumber\\
\overline{\mbox{{\bf 16}}}             &:& (+-++++--)
\ea
and their antiparticles, defined for the product of gamma matrices
$(\gl_5\gs_3\gs_6\gs_9\tgl_5\tgs_3\tgs_6\tgs_9)$. The second sector gives 4-0
generations with the chiralities
\ba
\label{eq:chiralitya}
\mbox{{\bf 16}}          &:& (++-----+),(+++--++-), \nonumber\\
                         & & (++-+-+-+),(++++-+++)
\ea
defined for the product of gamma matrices
$(\gl_5\gs_3\gs_5\gs_8\tgl_5\tgs_3\tgs_5\tgs_8)$. In order to give such a 4-1
model, we wish to construct an additional $\w_5$ which overlaps in such a way
that some of the old generations are projected out, and no new generations are
created. One way to do this is for $\w_5$ to give constraints that impose
$\gs_3\gs_6=-1$ in the first sector and $\gs_3\gs_8=-1$ in the second. A
suitable vector is
\be
\w_5 =
\lll (0)\com
\left(\h\com\h\com 0\right)\com
\left(\h\com\h\com 0\right)\com
\left(0 \com 0\com 0\right)\ccc
\left(0 \com 0\com\h\right)\com
\left(0 \com\h\com\h\right)\com
\left(0 \com\h\com 0\right)\com
(0)^5\com (\h)^4(0)^4 \rrr ,
\ee
with the new structure constants chosen to be all zero except $k_{25}=\h$. This
vector projects out the first generation of Eq.(\ref{eq:chirality}) and the
last two generations of Eq.(\ref{eq:chiralitya}). In addition, the overlap with
the $E_8^{\prime}$ degrees of freedom ensures that there are no new sectors
which could contain more generations. The gauge symmetry of the visible sector
is broken down to $SO(10)$, and with further vectors we could clearly arrange
to end up with smaller groups still. \hfill \break The second model is
initially a left--right symmetric 9-2 model and gives a 5-2 model. The
compactification vector is
\be
\w_4 =
\lll (0)\com
\left(0\com\frac{1}{5}\com\frac{4}{5}\right)\com
\left(\frac{1}{2}\com\frac{3}{5}\com\frac{9}{10}\right)\com
\left(\frac{3}{10}\com\frac{4}{5}\com\frac{9}{10}\right) \ccc
\left(0\com\frac{1}{5}\com\frac{4}{5}\right)\com
\left(\frac{1}{2}\com\frac{3}{5}\com\frac{9}{10}\right)\com
\left(\frac{3}{10}\com\frac{4}{5}\com\frac{9}{10}\right)\com
(0)^5\com (0)^8 \rrr .
\ee
Fermionic {\bf 16} representations come from the   $\overline{\wo + \w_1 +
\w_2}$ and $\overline{\wo + \w_1 + \w_2 + 5 \w_4}$ sectors. The first sector
has 5-0 generations, with the chiralities
\ba
\label{eq:chiralityb}
\mbox{{\bf 16}}          &:& (+--+---+),(+--+--+-),(+-+----+)  \nonumber\\
                            & & (+-+---+-),(++++-+++)
\ea
defined for the product of gamma matrices
$(\gl_5\gs_3\gs_6\gs_9\tgl_5\tgs_3\tgs_6\tgs_9)$ and the second sector has 4-2
generations with the chiralities
\ba
\label{eq:chiralityc}
\mbox{{\bf 16}}          &:& (+-------),(+--+---+), \nonumber\\
                            & & (+-+---+-),(+-++--++)  \nonumber\\
\overline{\mbox{{\bf 16}}}        &:& (+--+-++-),(+-++--++)
\ea
defined for the product of gamma matrices
$(\gl_5\gs_3\gs_4\gs_7\tgl_5\tgs_3\tgs_4\tgs_7)$. A 5-2 model is obtained by
adding the vector $\w_5$
\be
\w_5 =
\lll (0)\com
\left(0\com 0\com 0\right)\com
\left(0\com 0\com 0\right)\com
\left(0\com 0\com 0\right)\ccc
\left(0\com 0\com 0\right)\com
\left(0\com 0\com\h\right)\com
\left(0\com 0\com\h\right)\com
(0)^5\com (\h)^2(0)^6 \rrr ,
\ee
which projects out all the states in Eq.(\ref{eq:chiralityb}) which have
$\tgs_6\tgs_9 = -1$, and does not affect any of the states in
Eq.(\ref{eq:chiralityc}). Here we have to set all the new structure constants
to zero except $k_{54}=\frac{9}{10}$ and $k_{52}=\h$. \hfill \break We should
add that this second model may still allow the hope for a bosonisation into
a manifold (like discussed above), since $\bf{W_5}$ has nonzero entries in at
most one position of the left triplets.

Finally we address another aspect of the underlying $(2,2)$ models. Since we
have constructed $N=1$ supersymmetric space-time compactifications  with
maximal exceptional gauge groups, implying $N=2$ algebras on the world-sheet,
there must be the moduli fields associated to this structure. \hfill \break The
first set is obtained by acting with $G^+(\bar{z})$ on the left-handed (chiral)
{\bf 27} superfields, while the second set is obtained by acting with
$G^-(\bar{z})$ on the right-handed (antichiral)
$\overline{\mbox{{\bf 27}}}$
superfields. The explicit form of the algebra is only known for $Z_2$ boundary
conditions\footnote{ In this case the supercurrents are simply the linear
combination
\be
G^+ = -\sqrt{2} \sum_{j=1}^3 \psi_j\partial X_j,
\ee
\be
G^- = -\sqrt{2} \sum_{j=1}^3 \bar{\psi_j}\partial \bar{X_j},
\ee
The bosonisation procedure we described above may be used to give
\be
\sqrt{2}\partial z_j^* \equiv
i(:\overline{\rho}_j\rho_j:-i\overline{\sigma}_j\sigma_j:)
\ee
which is precisely the prescription given in ref.\cite{us} for an $N=2$ algebra
on the world-sheet.}, but nevertheless one is able to construct
the moduli by using the superpartner of the stress energy-tensor in the $N=1$
subalgebra of the $N=2$ algebra, that is known explicitly.
\begin{equation}
T_F(\bar{z}) = \frac{1}{\sqrt{2}} (G^+(\bar{z}) + G^-(\bar{z}))
             = i\sum_{i=1}^3
\tilde{\rho}^{i}\tilde{\sigma}^{i}\tilde{\psi}^{i} + \mbox{ h.c.}
\end{equation}
Using the fact that $G^-(\bar{z})$ vanishes on the left-handed superfields and
$G^+(\bar{z})$ vanishes on the rigth-handed ones, we may simply use
$T_F(\bar{z})$ to construct the moduli.

To demonstrate this explicitly let us give an example. Suppose that a
generation exists with a {\bf 10} of the form,
\be
b^{i}_{-v^i} |0\rangle\ox \bb^{j}_{-\vv^j} \bb^{10}_{-\h} |0\rangle
\ee
in a sector
\[
\aw = \lll (\h)\com
\left(v^1\com v^2\com v^3\right)\com
\left(v^4\com v^5\com v^6\right)\com
\left(v^7\com v^8\com v^9\right)\ccc
\left(\vv^1\com \vv^2\com \vv^3\right)\com
\left(\vv^4\com \vv^5\com \vv^6\right)\com
\left(\vv^7\com \vv^8\com \vv^9\right)\com
(\h)^5\com (\h)^8 \rrr ,
\]
and consider the singlet state which is generated from it by acting with the
$T_F(\bar{z})$, together with the removal of the $SO(10)$ excitation,
\be
b^{i}_{-v^i} |0\rangle\ox \dd^{k}_{-\vv^k} \dd^l_{-\vv^l} |0\rangle
\ee
where the indices $k$ and $l$ are in the same triplet as but not equal to $j$.
Clearly the $\w_1$ modular invariance condition is unchanged (see
ref.\cite{us}), but what about the conditions from the vectors overlapping?
Taking the constraint associated with $\w_n$
\be
\w_n\cdot\na=P_n\mod
\ee
where $P_n$ depends only on the sector and is the same for each state, we see
that for both the {\bf 10} and singlet to exist, we require
\be
\w_n^j+\w_n^{10}-\w_n^i=-\w_n^k-\w_n^l-\w_n^i\mod
\ee
But this is simply the triplet constraint and is therefore trivially satisfied.
Finally we have to show that the singlet is massless which also follows from
the triplet constraint since the vacuum energies for both states are the same.
In this way one can construct the moduli for all the models considered here by
acting with $T_F(\bar{z})$.

To conclude, we have given a classification of $(2,2)$ free field
compactifications that is expected to be exhaustive at the one twist level. The
fermionic and orbifold compactifications  overlap at least in the way predicted
by ref.\cite{PS89}. Further conclusions about the models in which the numbers
of  generations and antigenerations coincide with other orbifolds  have to be
postponed at the current state of knowledge. In particular, a conjecture such
as fermionic strings overlap with $Z_N$ orbifolds and Gepner models overlap
with $Z_N\times Z_M$ orbifolds, cannot be made unless the appearance of, for
example, the model with five generations and one antigenerations is explained.
\hfill \break Three generation models with no antigenerations have been found.
For the case of $(2,0)$ models, additional three generation models with
antigenerations have been given.

\vspace{1cm}
\noindent
{\bf\Large Acknowledgement} \hspace{0.3cm}
We would also like to thank the RAL computer division, especially Dick Roberts.
For discussions we would like to thank Michel Rausch de Traubenberg and in
particular Luis Iba\~{n}ez.

\bigskip
\bigskip

\section*{Table Captions}

\begin{description}
\item{\bf Table 1 }Supersymmetric (2,2) models in $D=4$. $n_{gen}$, $n_{agen}$
are the numbers of fundamental representations of $E_{6+X}$.
$n_g$ is the number of
bosons in $g$, and $n_s$ is the number of singlets. The gauge group is $g\ox
E_{6+X}\ox E^{\prime}_8$ where $2^{X} = N$. The models marked with a star may
only be generated with more complicated $\w_1$, $\w_2$ vectors. Here they have
an internal structure, $\left( 0   \frac{ 1}{ 6} \frac{ 2}{ 6}  \right)\left(0
\frac{ 1}{ 6}\frac{ 2}{ 6}\right)\left(0 \frac{ 1}{ 6}\frac{ 2}{ 6 }\right)$
\end{description}

                            %bottom

\clearpage
\begin{table}
\caption{\hspace{16 cm}}
\begin{tabular}{|r|r|r|r|r|r|r|r|}\hline\hline
 number & vector & N-SUSY     & $n_{gen}$&$n_{agen}$&$n_s$&$n_g$
        &$|\chi /2|$ \\ \hline
%%%put in here
   1&$\left(  0   0   0
\right)\left(  0   0   0
\right)\left(  0   0   0
\right)$&4&  1&  0&   -& 66& 0\\
   2&$\left(  0   0   0
\right)\left(\frac{ 8}{20}\frac{13}{20}\frac{19}{20
}\right)\left(\frac{ 7}{20}\frac{14}{20}\frac{19}{20
}\right)$&2& 10&  0&  36& 13& 0\\
   4&$\left(  0   0   0
\right)\left(  0 \frac{ 1}{16}\frac{15}{16
}\right)\left(\frac{ 6}{16}\frac{11}{16}\frac{15}{16
}\right)$&2&  9&  0&  63& 19& 0\\
   3&$\left(  0   0   0
\right)\left(\frac{12}{20}\frac{13}{20}\frac{15}{20
}\right)\left(\frac{11}{20}\frac{14}{20}\frac{15}{20
}\right)$&2&  8&  0&  16& 11& 0\\
   5&$\left(  0   0   0
\right)\left( 0 \frac{ 1}{ 8}\frac{ 7}{ 8
}\right)\left(\frac{ 3}{ 8}\frac{ 6}{ 8}\frac{ 7}{ 8
}\right)$&2&  7&  0&  56& 19& 0\\
   6&$\left( 0  0  0
\right)\left( 0 \frac{ 1}{ 2}\frac{ 1}{ 2
}\right)\left( 0 \frac{ 1}{ 2}\frac{ 1}{ 2
}\right)$&2&  6&  0&  96& 37& 0\\
   7&$\left( 0  0  0
\right)\left(\frac{ 3}{ 6}\frac{ 4}{ 6}\frac{ 5}{ 6
}\right)\left(\frac{ 3}{ 6}\frac{ 4}{ 6}\frac{ 5}{ 6
}\right)$&2&  5&  0&  24& 17& 0\\
   8&$\left( 0  0  0
\right)\left( 0 \frac{ 1}{ 4}\frac{ 3}{ 4
}\right)\left( 0 \frac{ 1}{ 4}\frac{ 3}{ 4
}\right)$&2&  4&  0&  48& 33& 0\\
   9&$\left( 0  0  0
\right)\left( 0 \frac{ 1}{ 6}\frac{ 5}{ 6
}\right)\left(\frac{ 3}{ 6}\frac{ 4}{ 6}\frac{ 5}{ 6
}\right)$&2&  3&  0&  27& 19& 0\\
  10&$\left( 0  0  0
\right)\left( 0 \frac{ 2}{ 4}\frac{ 2}{ 4
}\right)\left(\frac{ 1}{ 4}\frac{ 1}{ 4}\frac{ 2}{ 4
}\right)$&2&  2&  0&  14& 23& 0\\
  11&$\left( 0  0  0
\right)\left( 0 \frac{ 2}{ 6}\frac{ 4}{ 6
}\right)\left( 0 \frac{ 2}{ 6}\frac{ 4}{ 6
}\right)$&2&  1&  0&  18& 33& 0\\
  12&$\left( 0 \frac{ 2}{16}\frac{14}{16
}\right)\left( 0 \frac{ 3}{16}\frac{13}{16
}\right)\left(\frac{ 2}{16}\frac{ 3}{16}\frac{11}{16
}\right)$&1& 51&  0& 212& 16&51\\
  13&$\left( 0 \frac{ 2}{20}\frac{18}{20
}\right)\left( 0 \frac{ 3}{20}\frac{17}{20
}\right)\left(\frac{ 3}{20}\frac{18}{20}\frac{19}{20
}\right)$&1& 46&  0& 204& 16&46\\
  14&$\left( 0 \frac{ 2}{ 8}\frac{ 6}{ 8
}\right)\left( 0 \frac{ 1}{ 8}\frac{ 7}{ 8
}\right)\left(\frac{ 2}{ 8}\frac{ 7}{ 8}\frac{ 7}{ 8
}\right)$&1& 41&  1& 264& 18&40\\
  15&$\left( 0 \frac{ 4}{16}\frac{12}{16
}\right)\left( 0 \frac{ 3}{16}\frac{13}{16
}\right)\left(\frac{ 3}{16}\frac{14}{16}\frac{15}{16
}\right)$&1& 39&  0& 154& 14&39\\
  16&$\left( 0 \frac{ 2}{16}\frac{14}{16
}\right)\left( 0 \frac{ 3}{16}\frac{13}{16
}\right)\left(\frac{ 2}{16}\frac{15}{16}\frac{15}{16
}\right)$&1& 37&  1& 214& 14&36\\
  17&$\left( 0 \frac{ 2}{16}\frac{14}{16
}\right)\left(\frac{ 8}{16}\frac{11}{16}\frac{13}{16
}\right)\left(\frac{10}{16}\frac{11}{16}\frac{11}{16
}\right)$&1& 35&  0& 174& 10&35\\
  18&$\left( 0 \frac{ 2}{12}\frac{10}{12
}\right)\left( 0 \frac{ 1}{12}\frac{11}{12
}\right)\left(\frac{ 3}{12}\frac{10}{12}\frac{11}{12
}\right)$&1& 35&  1& 208& 16&34\\
  19&$\left( 0 \frac{ 4}{16}\frac{12}{16
}\right)\left(\frac{ 6}{16}\frac{13}{16}\frac{13}{16
}\right)\left(\frac{ 3}{16}\frac{ 4}{16}\frac{ 9}{16
}\right)$&1& 33&  0& 152& 12&33\\
  20&$\left( 0 \frac{ 4}{12}\frac{ 8}{12
}\right)\left( 0 \frac{ 1}{12}\frac{11}{12
}\right)\left(\frac{ 1}{12}\frac{ 2}{12}\frac{ 9}{12
}\right)$&1& 31&  0& 147& 14&31\\
  21&$\left( 0 \frac{ 2}{16}\frac{14}{16
}\right)\left( 0 \frac{ 3}{16}\frac{13}{16
}\right)\left(\frac{10}{16}\frac{11}{16}\frac{11}{16
}\right)$&1& 31&  1& 134& 12&30\\
  22&$\left( 0 \frac{ 4}{20}\frac{16}{20
}\right)\left(\frac{ 2}{20}\frac{19}{20}\frac{19}{20
}\right)\left(\frac{ 1}{20}\frac{ 4}{20}\frac{15}{20
}\right)$&1& 30&  0& 118& 12&30\\
  23&$\left( 0 \frac{ 4}{16}\frac{12}{16
}\right)\left(\frac{ 7}{16}\frac{12}{16}\frac{13}{16
}\right)\left(\frac{ 1}{16}\frac{ 6}{16}\frac{ 9}{16
}\right)$&1& 29&  0& 118& 10&29\\
  24&$\left( 0 \frac{ 2}{16}\frac{14}{16
}\right)\left(\frac{ 4}{16}\frac{ 5}{16}\frac{ 7}{16
}\right)\left(\frac{ 3}{16}\frac{ 6}{16}\frac{ 7}{16
}\right)$&1& 29&  1& 124& 10&28\\
  25&$\left( 0 \frac{ 4}{16}\frac{12}{16
}\right)\left( 0 \frac{ 1}{16}\frac{15}{16
}\right)\left(\frac{ 9}{16}\frac{10}{16}\frac{13}{16
}\right)$&1& 27&  0&  96& 12&27\\
  26&$\left( 0 \frac{ 8}{16}\frac{ 8}{16
}\right)\left(\frac{ 1}{16}\frac{ 2}{16}\frac{13}{16
}\right)\left( 0 \frac{ 5}{16}\frac{11}{16
}\right)$&1& 27&  1& 128& 12&26\\
  27&$\left( 0 \frac{ 1}{10}\frac{ 9}{10
}\right)\left( 0 \frac{ 1}{10}\frac{ 9}{10
}\right)\left(\frac{ 1}{10}\frac{ 1}{10}\frac{ 8}{10
}\right)$&1& 26&  0& 204& 26&26\\
  28&$\left( 0 \frac{ 5}{20}\frac{15}{20
}\right)\left(\frac{ 4}{20}\frac{ 6}{20}\frac{10}{20
}\right)\left(\frac{ 2}{20}\frac{ 3}{20}\frac{15}{20
}\right)$&1& 26&  1&  77& 10&25\\
  29&$\left( 0 \frac{ 4}{16}\frac{12}{16
}\right)\left( 0 \frac{ 1}{16}\frac{15}{16
}\right)\left(\frac{ 1}{16}\frac{ 2}{16}\frac{13}{16
}\right)$&1& 25&  0& 120& 14&25\\
  30&$\left( 0 \frac{ 2}{ 4}\frac{ 2}{ 4
}\right)\left( 0 \frac{ 1}{ 4}\frac{ 3}{ 4
}\right)\left(\frac{ 2}{ 4}\frac{ 3}{ 4}\frac{ 3}{ 4
}\right)$&1& 27&  3& 270& 20&24\\
  31&$\left( 0 \frac{ 4}{ 8}\frac{ 4}{ 8
}\right)\left( 0 \frac{ 1}{ 8}\frac{ 7}{ 8
}\right)\left(\frac{ 1}{ 8}\frac{ 2}{ 8}\frac{ 5}{ 8
}\right)$&1& 25&  1& 168& 14&24\\
  32&$\left( 0 \frac{ 4}{12}\frac{ 8}{12
}\right)\left(\frac{ 6}{12}\frac{ 7}{12}\frac{11}{12
}\right)\left(\frac{ 7}{12}\frac{ 8}{12}\frac{ 9}{12
}\right)$&1& 24&  0& 122& 12&24\\
  33&$\left( 0 \frac{ 4}{12}\frac{ 8}{12
}\right)\left(\frac{ 6}{12}\frac{ 7}{12}\frac{11}{12
}\right)\left(\frac{ 3}{12}\frac{ 4}{12}\frac{ 5}{12
}\right)$&1& 24&  1& 124& 10&23\\
  34&$\left( 0 \frac{ 3}{12}\frac{ 9}{12
}\right)\left( 0 \frac{ 2}{12}\frac{10}{12
}\right)\left(\frac{ 3}{12}\frac{10}{12}\frac{11}{12
}\right)$&1& 23&  0& 102& 16&23\\
  35&$\left( 0 \frac{ 2}{ 8}\frac{ 6}{ 8
}\right)\left( 0 \frac{ 1}{ 8}\frac{ 7}{ 8
}\right)\left(\frac{ 1}{ 8}\frac{ 2}{ 8}\frac{ 5}{ 8
}\right)$&1& 23&  1& 134& 16&22\\
  36&$\left( 0 \frac{ 2}{ 6}\frac{ 4}{ 6
}\right)\left(\frac{ 3}{ 6}\frac{ 4}{ 6}\frac{ 5}{ 6
}\right)\left(\frac{ 3}{ 6}\frac{ 4}{ 6}\frac{ 5}{ 6
}\right)$&1& 22&  0& 114& 20&22\\
  37&$\left( 0 \frac{ 4}{16}\frac{12}{16
}\right)\left(\frac{ 8}{16}\frac{11}{16}\frac{13}{16
}\right)\left(\frac{ 2}{16}\frac{ 5}{16}\frac{ 9}{16
}\right)$&1& 21&  0&  92& 10&21\\
  38&$\left( 0 \frac{ 2}{ 6}\frac{ 4}{ 6
}\right)\left( 0 \frac{ 1}{ 6}\frac{ 5}{ 6
}\right)\left( 0 \frac{ 1}{ 6}\frac{ 5}{ 6
}\right)$&1& 21&  1& 162& 24&20\\
  39&$\left( 0 \frac{ 4}{12}\frac{ 8}{12
}\right)\left(\frac{ 1}{12}\frac{ 3}{12}\frac{ 8}{12
}\right)\left(\frac{ 1}{12}\frac{ 3}{12}\frac{ 8}{12
}\right)$&1& 20&  0&  60& 18&20\\
  40&$\left( 0 \frac{ 3}{12}\frac{ 9}{12
}\right)\left(\frac{ 4}{12}\frac{10}{12}\frac{10}{12
}\right)\left(\frac{ 3}{12}\frac{10}{12}\frac{11}{12
}\right)$&1& 19&  0&  84& 12&19\\
  41&$\left( 0 \frac{ 2}{ 8}\frac{ 6}{ 8
}\right)\left(\frac{ 4}{ 8}\frac{ 5}{ 8}\frac{ 7}{ 8
}\right)\left(\frac{ 2}{ 8}\frac{ 3}{ 8}\frac{ 3}{ 8
}\right)$&1& 19&  1& 130& 12&18\\
  42&$\left( 0 \frac{ 2}{ 6}\frac{ 4}{ 6
}\right)\left( 0 \frac{ 2}{ 6}\frac{ 4}{ 6
}\right)\left( 0 \frac{ 2}{ 6}\frac{ 4}{ 6
}\right)$&1& 18&  0& 126& 32&18\\
  43&$\left( 0 \frac{ 2}{18}\frac{16}{18
}\right)\left(\frac{ 3}{18}\frac{ 4}{18}\frac{11}{18
}\right)\left(\frac{ 2}{18}\frac{ 7}{18}\frac{ 9}{18
}\right)$&1& 18&  1&  84& 10&17\\
  44&$\left( 0 \frac{ 3}{12}\frac{ 9}{12
}\right)\left(\frac{ 6}{12}\frac{ 8}{12}\frac{10}{12
}\right)\left(\frac{ 3}{12}\frac{10}{12}\frac{11}{12
}\right)$&1& 17&  0&  90& 10&17\\
\hline\hline
\end{tabular}
\end{table}
\clearpage
\begin{table}
\begin{tabular}{|r|r|r|r|r|r|r|r|}\hline\hline
 number & vector & N-SUSY     & $n_{gen}$&$n_{agen}$&$n_s$&$n_g$
        &$|\chi /2|$ \\ \hline
  45&$\left( 0 \frac{ 6}{12}\frac{ 6}{12
}\right)\left(\frac{ 1}{12}\frac{ 4}{12}\frac{ 7}{12
}\right)\left(\frac{ 6}{12}\frac{ 7}{12}\frac{11}{12
}\right)$&1& 19&  3& 132& 12&16\\
  46&$\left( 0 \frac{ 2}{10}\frac{ 8}{10
}\right)\left(\frac{ 2}{10}\frac{ 3}{10}\frac{ 5}{10
}\right)\left(\frac{ 6}{10}\frac{ 7}{10}\frac{ 7}{10
}\right)$&1& 17&  1&  94& 12&16\\
  47&$\left( 0 \frac{ 2}{10}\frac{ 8}{10
}\right)\left(\frac{ 1}{10}\frac{ 3}{10}\frac{ 6}{10
}\right)\left(\frac{ 1}{10}\frac{ 3}{10}\frac{ 6}{10
}\right)$&1& 16&  0&  61& 14&16\\
  48&$\left( 0 \frac{ 2}{12}\frac{10}{12
}\right)\left(\frac{ 6}{12}\frac{ 7}{12}\frac{11}{12
}\right)\left(\frac{ 3}{12}\frac{10}{12}\frac{11}{12
}\right)$&1& 16&  1&  58& 12&15\\
  49&$\left( 0 \frac{ 1}{16}\frac{15}{16
}\right)\left(\frac{ 4}{16}\frac{14}{16}\frac{14}{16
}\right)\left(\frac{ 9}{16}\frac{10}{16}\frac{13}{16
}\right)$&1& 15&  0&  62&  8&15\\
  50&$\left( 0 \frac{ 6}{12}\frac{ 6}{12
}\right)\left(\frac{ 1}{12}\frac{ 2}{12}\frac{ 9}{12
}\right)\left(\frac{ 7}{12}\frac{ 8}{12}\frac{ 9}{12
}\right)$&1& 17&  3&  90& 10&14\\
  51&$\left( 0 \frac{ 2}{ 6}\frac{ 4}{ 6
}\right)\left( 0 \frac{ 1}{ 6}\frac{ 5}{ 6
}\right)\left(\frac{ 3}{ 6}\frac{ 4}{ 6}\frac{ 5}{ 6
}\right)$&1& 16&  2& 134& 16&14\\
  52&$\left( 0 \frac{ 6}{12}\frac{ 6}{12
}\right)\left(\frac{ 1}{12}\frac{ 4}{12}\frac{ 7}{12
}\right)\left(\frac{ 3}{12}\frac{10}{12}\frac{11}{12
}\right)$&1& 15&  1&  90&  8&14\\
  53&$\left( 0 \frac{ 2}{10}\frac{ 8}{10
}\right)\left(\frac{ 5}{10}\frac{ 7}{10}\frac{ 8}{10
}\right)\left(\frac{ 2}{10}\frac{ 2}{10}\frac{ 6}{10
}\right)$&1& 14&  0&  74& 16&14\\
  54&$\left( 0 \frac{ 4}{12}\frac{ 8}{12
}\right)\left( 0 \frac{ 1}{12}\frac{11}{12
}\right)\left(\frac{ 3}{12}\frac{ 4}{12}\frac{ 5}{12
}\right)$&1& 14&  1&  84& 14&13\\
  55&$\left( 0 \frac{ 1}{ 6}\frac{ 5}{ 6
}\right)\left( 0 \frac{ 1}{ 6}\frac{ 5}{ 6
}\right)\left(\frac{ 1}{ 6}\frac{ 1}{ 6}\frac{ 4}{ 6
}\right)$&1& 13&  0& 102& 26&13\\
  56&$\left( 0 \frac{ 1}{18}\frac{17}{18
}\right)\left(\frac{ 3}{18}\frac{ 7}{18}\frac{ 8}{18
}\right)\left(\frac{ 1}{18}\frac{ 8}{18}\frac{ 9}{18
}\right)$&1& 15&  3&  89& 10&12\\
  57&$\left( 0 \frac{ 1}{12}\frac{11}{12
}\right)\left(\frac{ 3}{12}\frac{ 4}{12}\frac{ 5}{12
}\right)\left(\frac{ 1}{12}\frac{ 3}{12}\frac{ 8}{12
}\right)$&1& 14&  2&  70& 12&12\\
  58&$\left( 0 \frac{ 1}{ 4}\frac{ 3}{ 4
}\right)\left( 0 \frac{ 1}{ 4}\frac{ 3}{ 4
}\right)\left(\frac{ 1}{ 4}\frac{ 1}{ 4}\frac{ 2}{ 4
}\right)$&1& 13&  1& 120& 26&12\\
  59&$\left( 0 \frac{ 2}{10}\frac{ 8}{10
}\right)\left(\frac{ 1}{10}\frac{ 3}{10}\frac{ 6}{10
}\right)\left(\frac{ 6}{10}\frac{ 6}{10}\frac{ 8}{10
}\right)$&1& 12&  0&  71& 12&12\\
  60&$\left( 0 \frac{ 1}{18}\frac{17}{18
}\right)\left(\frac{ 1}{18}\frac{ 2}{18}\frac{15}{18
}\right)\left(\frac{ 5}{18}\frac{15}{18}\frac{16}{18
}\right)$&1& 13&  2&  83& 10&11\\
  61&$\left( 0 \frac{ 4}{12}\frac{ 8}{12
}\right)\left(\frac{ 6}{12}\frac{ 7}{12}\frac{11}{12
}\right)\left(\frac{ 2}{12}\frac{ 3}{12}\frac{ 7}{12
}\right)$&1& 12&  1&  66&  8&11\\
  62&$\left( 0 \frac{ 1}{ 8}\frac{ 7}{ 8
}\right)\left( 0 \frac{ 3}{ 8}\frac{ 5}{ 8
}\right)\left(\frac{ 1}{ 8}\frac{ 1}{ 8}\frac{ 6}{ 8
}\right)$&1& 11&  0&  56& 18&11\\
  63&$\left( 0 \frac{ 2}{10}\frac{ 8}{10
}\right)\left(\frac{ 2}{10}\frac{ 3}{10}\frac{ 5}{10
}\right)\left(\frac{ 1}{10}\frac{ 2}{10}\frac{ 7}{10
}\right)$&1& 12&  2&  76& 14&10\\
%\hline\hline
%\end{tabular}
%\end{table}
%\clearpage
%\begin{table}
%\begin{tabular}{|r|r|r|r|r|r|r|r|}\hline\hline
% number & vector & N-SUSY     & $n_{gen}$&$n_{agen}$&$n_s$&$n_g$
%        &$|\chi /2|$ \\ \hline
  64&$\left( 0 \frac{ 3}{ 6}\frac{ 3}{ 6
}\right)\left( 0 \frac{ 2}{ 6}\frac{ 4}{ 6
}\right)\left(\frac{ 3}{ 6}\frac{ 4}{ 6}\frac{ 5}{ 6
}\right)$&1& 11&  1&  84& 18&10\\
  65&$\left( 0 \frac{ 2}{10}\frac{ 8}{10
}\right)\left( 0 \frac{ 2}{10}\frac{ 8}{10
}\right)\left(\frac{ 2}{10}\frac{ 2}{10}\frac{ 6}{10
}\right)$&1& 10&  0&  80& 26&10\\
  66&$\left( 0 \frac{ 3}{12}\frac{ 9}{12
}\right)\left(\frac{ 2}{12}\frac{ 5}{12}\frac{ 5}{12
}\right)\left(\frac{ 1}{12}\frac{ 1}{12}\frac{10}{12
}\right)$&1& 11&  2&  59& 10& 9\\
  67&$\left( 0 \frac{ 2}{ 6}\frac{ 4}{ 6
}\right)\left( 0 \frac{ 2}{ 6}\frac{ 4}{ 6
}\right)\left(\frac{ 2}{ 6}\frac{ 2}{ 6}\frac{ 2}{ 6
}\right)$&1&  9&  0&  66& 30& 9\\
  68&$\left( 0 \frac{ 1}{16}\frac{15}{16
}\right)\left(\frac{ 4}{16}\frac{ 4}{16}\frac{ 8}{16
}\right)\left(\frac{ 3}{16}\frac{ 6}{16}\frac{ 7}{16
}\right)$&1& 15&  7& 108& 10& 8\\
  69&$\left( 0 \frac{ 2}{ 4}\frac{ 2}{ 4
}\right)\left( 0 \frac{ 1}{ 4}\frac{ 3}{ 4
}\right)\left( 0 \frac{ 1}{ 4}\frac{ 3}{ 4
}\right)$&1& 11&  3& 122& 24& 8\\
  70&$\left( 0 \frac{ 4}{ 8}\frac{ 4}{ 8
}\right)\left( 0 \frac{ 3}{ 8}\frac{ 5}{ 8
}\right)\left(\frac{ 4}{ 8}\frac{ 5}{ 8}\frac{ 7}{ 8
}\right)$&1&  9&  1&  72& 18& 8\\
  71&$\left( 0 \frac{ 2}{14}\frac{12}{14
}\right)\left(\frac{ 2}{14}\frac{ 4}{14}\frac{ 8}{14
}\right)\left(\frac{ 2}{14}\frac{ 6}{14}\frac{ 6}{14
}\right)$&1&  8&  0&  55& 16& 8\\
  72&$\left( 0 \frac{ 2}{ 6}\frac{ 4}{ 6
}\right)\left(\frac{ 3}{ 6}\frac{ 4}{ 6}\frac{ 5}{ 6
}\right)\left(\frac{ 2}{ 6}\frac{ 5}{ 6}\frac{ 5}{ 6
}\right)$&1&  9&  2&  81& 16& 7\\
  73&$\left( 0 \frac{ 2}{10}\frac{ 8}{10
}\right)\left(\frac{ 5}{10}\frac{ 6}{10}\frac{ 9}{10
}\right)\left(\frac{ 1}{10}\frac{ 2}{10}\frac{ 7}{10
}\right)$&1&  7&  0&  48& 10& 7\\
  74&$\left( 0 \frac{ 2}{16}\frac{14}{16
}\right)\left(\frac{ 1}{16}\frac{ 7}{16}\frac{ 8}{16
}\right)\left(\frac{ 5}{16}\frac{ 5}{16}\frac{ 6}{16
}\right)$&1& 11&  5&  72& 10& 6\\
  75&$\left( 0 \frac{10}{20}\frac{10}{20
}\right)\left(\frac{ 6}{20}\frac{15}{20}\frac{19}{20
}\right)\left(\frac{ 5}{20}\frac{ 6}{20}\frac{ 9}{20
}\right)$&1&  9&  3&  40& 12& 6\\
  76&$\left( 0 \frac{ 2}{ 8}\frac{ 6}{ 8
}\right)\left( 0 \frac{ 1}{ 8}\frac{ 7}{ 8
}\right)\left( 0 \frac{ 3}{ 8}\frac{ 5}{ 8
}\right)$&1&  7&  1&  60& 20& 6\\
  77&$\left( 0 \frac{ 2}{14}\frac{12}{14
}\right)\left( 0 \frac{ 4}{14}\frac{10}{14
}\right)\left( 0 \frac{ 6}{14}\frac{ 8}{14
}\right)$&1&  6&  0&  54& 20& 6\\
  78&$\left( 0 \frac{ 1}{ 8}\frac{ 7}{ 8
}\right)\left(\frac{ 2}{ 8}\frac{ 3}{ 8}\frac{ 3}{ 8
}\right)\left(\frac{ 1}{ 8}\frac{ 1}{ 8}\frac{ 6}{ 8
}\right)$&1&  7&  2&  60& 14& 5\\
  79&$\left( 0 \frac{ 2}{10}\frac{ 8}{10
}\right)\left( 0 \frac{ 2}{10}\frac{ 8}{10
}\right)\left( 0 \frac{ 4}{10}\frac{ 6}{10
}\right)$&1&  5&  0&  39& 24& 5\\
  80&$\left( 0 \frac{ 1}{ 4}\frac{ 3}{ 4
}\right)\left(\frac{ 2}{ 4}\frac{ 3}{ 4}\frac{ 3}{ 4
}\right)\left(\frac{ 1}{ 4}\frac{ 1}{ 4}\frac{ 2}{ 4
}\right)$&1&  9&  5& 124& 22& 4\\
  81&$\left( 0 \frac{ 3}{ 6}\frac{ 3}{ 6
}\right)\left( 0 \frac{ 2}{ 6}\frac{ 4}{ 6
}\right)\left( 0 \frac{ 1}{ 6}\frac{ 5}{ 6
}\right)$&1&  7&  3&  84& 20& 4\\
  82&$\left( 0 \frac{ 4}{ 8}\frac{ 4}{ 8
}\right)\left( 0 \frac{ 2}{ 8}\frac{ 6}{ 8
}\right)\left(\frac{ 1}{ 8}\frac{ 1}{ 8}\frac{ 6}{ 8
}\right)$&1&  5&  1&  26& 16& 4\\
  83&$\left( 0 \frac{ 2}{14}\frac{12}{14
}\right)\left( 0 \frac{ 4}{14}\frac{10}{14
}\right)\left(\frac{ 2}{14}\frac{ 4}{14}\frac{ 8}{14
}\right)$&1&  3&  0&  29& 16& 3\\
  84&$\left( 0 \frac{ 2}{18}\frac{16}{18
}\right)\left(\frac{ 2}{18}\frac{ 7}{18}\frac{ 9}{18
}\right)\left(\frac{ 5}{18}\frac{ 6}{18}\frac{ 7}{18
}\right)$&1&  9&  7& 108& 10& 2\\
  85&$\left( 0 \frac{ 1}{18}\frac{17}{18
}\right)\left(\frac{ 1}{18}\frac{ 8}{18}\frac{ 9}{18
}\right)\left(\frac{ 4}{18}\frac{ 6}{18}\frac{ 8}{18
}\right)$&1&  7&  5&  61& 10& 2\\
  $\star$ 86&$\left( 0  0  0
\right)\left( 0 \frac{ 3}{12}\frac{ 9}{12
}\right)\left(\frac{ 2}{12}\frac{11}{12}\frac{11}{12
}\right)$&1&  6&  4& 144& 14& 2\\
  87&$\left( 0 \frac{ 1}{ 6}\frac{ 5}{ 6
}\right)\left(\frac{ 2}{ 6}\frac{ 2}{ 6}\frac{ 2}{ 6
}\right)\left(\frac{ 1}{ 6}\frac{ 2}{ 6}\frac{ 3}{ 6
}\right)$&1&  5&  3&  84& 16& 2\\
  88&$\left( 0 \frac{ 6}{12}\frac{ 6}{12
}\right)\left( 0 \frac{ 4}{12}\frac{ 8}{12
}\right)\left(\frac{ 1}{12}\frac{ 1}{12}\frac{10}{12
}\right)$&1&  3&  1&  18& 14& 2\\
\hline\hline
\end{tabular}
\end{table}
\clearpage
\begin{table}
\begin{tabular}{|r|r|r|r|r|r|r|r|}\hline\hline
 number & vector & N-SUSY     & $n_{gen}$&$n_{agen}$&$n_s$&$n_g$
        &$|\chi /2|$ \\ \hline
  $\star$ 89&$\left( 0  0  0
\right)\left(\frac{ 4}{18}\frac{15}{18}\frac{17}{18
}\right)\left(\frac{ 2}{18}\frac{ 3}{18}\frac{13}{18
}\right)$&1&  2&  0&  31& 10& 2\\
  $\star$ 90&$\left( 0  0  0
\right)\left(\frac{ 1}{18}\frac{ 7}{18}\frac{10}{18
}\right)\left(\frac{ 1}{18}\frac{ 1}{18}\frac{16}{18
}\right)$&1&  2&  1&  21& 10& 1\\
  $\star$ 91&$\left( 0  0  0
\right)\left( 0 \frac{ 5}{18}\frac{13}{18
}\right)\left(\frac{ 5}{18}\frac{15}{18}\frac{16}{18
}\right)$&1&  1&  0&  12& 20& 1\\
  92&$\left( 0 \frac{ 1}{ 8}\frac{ 7}{ 8
}\right)\left(\frac{ 2}{ 8}\frac{ 2}{ 8}\frac{ 4}{ 8
}\right)\left(\frac{ 2}{ 8}\frac{ 3}{ 8}\frac{ 3}{ 8
}\right)$&1& 11& 11& 172& 14& 0\\
  93&$\left( 0 \frac{ 1}{12}\frac{11}{12
}\right)\left(\frac{ 3}{12}\frac{ 4}{12}\frac{ 5}{12
}\right)\left(\frac{ 1}{12}\frac{ 5}{12}\frac{ 6}{12
}\right)$&1&  5&  5&  57& 10& 0\\
  94&$\left( 0 \frac{ 1}{10}\frac{ 9}{10
}\right)\left(\frac{ 2}{10}\frac{ 4}{10}\frac{ 4}{10
}\right)\left(\frac{ 2}{10}\frac{ 3}{10}\frac{ 5}{10
}\right)$&1&  3&  3&  50& 10& 0\\
  $\star$ 95&$\left( 0  0  0
\right)\left( 0 \frac{ 2}{18}\frac{16}{18
}\right)\left(\frac{ 2}{18}\frac{ 6}{18}\frac{10}{18
}\right)$&1&  1&  1&  23& 14& 0\\
\hline\hline
\end{tabular}
\vspace{17.5cm}
\end{table}

\end{document}